\documentstyle[12pt,aaspp4,psfig]{article}

\lefthead{Le F\`evre et al.}
\righthead{}

\begin{document}

\title{Clustering around the radio-galaxy MRC0316-257 at z=3.14}

\author{O. Le F\`evre\altaffilmark{1}, J.M. Deltorn}
\affil{DAEC, Observatoire de Paris-Meudon, 92195 Meudon Cedex, France}

\author{D. Crampton}
\affil{Dominion Astrophysical Observatory, National Research Council of Canada, R.R. 5 Victoria, B.C., V8X4M6, Canada}

\and

\author{M. Dickinson}
\affil{Space Telescope Science Institute, Baltimore, MD 21218, USA}


\altaffiltext{1}{Visiting Astronomer, Canada France Hawaii Telescope. 
CFHT is operated by the National Research Council of Canada, the Centre National de la Recherche Scientifique of France, and the University of Hawaii.} 


\begin{abstract}
We report here the spectroscopic identification of galaxies in the neighborhood of the radio-galaxy MRC0316-257,  at a redshift $z\sim3.14$. Candidate cluster galaxies were selected from deep V and I images combined with narrow band imaging at the wavelength of redshifted Ly$\alpha$. Follow-up multi-slit spectroscopy has allowed confirmation of the redshift of the radio-galaxy, $z=3.1420\pm0.0020$, and identification of two associated galaxies at redshifts $z=3.1378\pm0.0028$ and $z=3.1351\pm0.0028$ respectively. The first galaxy is 0.3 $h_{50}^{-1}$ Mpc from the radio-galaxy,  is resolved with an intrinsic size $11.6\pm h_{50}^{-1}$ kpc, and shows $Ly\alpha$ in emission with rest $W_{Ly\alpha}=55\pm14$\AA. In addition, its extremely blue $V-I$ color might possibly indicate a proto-galaxy forming a first generation of stars in a low dust medium. The second galaxy is 1.3 $h_{50}^{-1}$ Mpc away from the radio-galaxy, is marginally resolved and, in addition to Ly$\alpha$ in emission, shows CIV in emission with a broad component indicating the contribution from an AGN. The comoving density of galaxies with $V<23.8$ and a $Ly\alpha$ flux $>10^{-16}$ ergcm$^{-2}$sec$^{-1}$ in the vicinity of MRC0316-257 is $\sim2.5\times10^{-3}h^3_{50}Mpc^{-3}$, significantly higher than the expected background density of field galaxies with similar properties, and might indicate a rich cluster or proto-cluster environment. These observations indicate that the environment of high redshift radio-galaxies may provide significant numbers of galaxies from which to study the early stages of cluster formation and galaxy evolution.
\end{abstract}




%

\section{Introduction}

The evolution of clusters of galaxies has important implications to our knowledge of the fundamental cosmological parameters. As they are the most massive gravitationally bound systems in the universe, the evolution of clusters is highly sensitive to the physics of cosmic structure formation and the values of the fundamental cosmological parameters. Moreover, clusters are valuable laboratories to study the evolution of many galaxies at a common redshift.

Although a number of  galaxies with luminosities $\sim L^*$ are now observed out to redshifts  $\sim 3-4$ (\cite{stei,fon}), very few candidate clusters of galaxies have been spectroscopically confirmed at redshifts near or above unity (\cite{gia,olf1,dic2,pas,fran}). At these very high redshifts, the prevalence of clusters, proto-clusters, or other large scale structures in the galaxy distribution is as yet unknown, and the evolution of large scale structures may well be at a critical stage (\cite{pee,evr,frenk}), where observations can directly constrain cosmological models. 

At intermediate redshifts, powerful radio-galaxies are  frequently located in rich clusters (\cite{yee,hill,dic}), therefore, a possible search strategy is  to look for clusters around known powerful radio-galaxies. 
As part of a program to identify clusters and study their properties at redshifts larger than $z\sim0.8$, we  report here the discovery of galaxies within $\sim1 h_{50}^{-1}$ Mpc and 1000 kms$^{-1}$ from the 1Jy radio galaxy MRC0316-257 (\cite{Mcc}), at $z=3.14$. 

$H_0=50$ kms$^{-1}$Mpc$^{-1}$ and $q_0=0.5$ are used throughout this letter.

\section{Observations}

We have observed a $9.2\arcmin\times8.5\arcmin$ field around MRC0316-257 with the Multi Object imaging Spectrograph (MOS) at CFHT (\cite{olf2}). Deep broad band V and I images as well as narrow band images in a filter with central wavelength 5007\AA ~and bandwidth 96\AA, containing the $Ly\alpha$ line redshifted to $z\sim3.14$, were obtained the night of 20 December 1995. Several images were obtained for each filter, after small telescope offsets were performed, to improve the flat fielding accuracy. Total integration times were 2700, 1200, 5100 seconds in V, I and the 5007\AA ~filters, respectively. Standard image processing was performed the next day, and the V and 5007\AA ~images were blinked visually to provide a list of cluster candidates for subsequent multi-slit spectroscopy during the nights 21-24 December 1995. The narrow band 5007\AA ~images can be  expected to identify galaxies with $Ly\alpha$ in a redshift range $z=3.081-3.160$,  foreground galaxies with [OII]3727\AA ~in emission with a redshift $z=0.330-0.356$, or the rarer occurence of galaxies with [OIII]5007\AA ~with $z\sim0$. 

The narrow band image is compared to the V band image of the MRC0316-257 field in 
Figure 1 (plate 1). Two galaxies exhibit strong excess emission in the narrow band filter, and were subsequently observed spectroscopically, together with more marginal candidates.
Multi-slit masks were prepared with the highest priority given to the cluster candidates as defined above, and other slits were placed on galaxies brighter than $V\sim23.8$ when no other candidate was available. The observations of about 30 galaxies per mask were performed with the O300 grism, with 300 l/mm and peak transmission at $\sim$5900\AA, ~and slits $1.75\arcsec$ wide, providing a spectral range of 4500-9000\AA ~and a resolution of $20$\AA. Three masks have received cummulated exposure times of 19800, 8700, and 8700 seconds. Data reduction of the spectra in the first mask obtained the night of 21 December 1995 was performed the following afternoon, and confirmed that these candidate galaxies had emission lines at $\sim$5030\AA. These galaxies were subsequently reobserved in the following multi-slit mask for a total exposure time of 28500 sec. 

Final data reduction was performed with the MULTIRED package implemented under IRAF, and redshifts were measured and assigned classes as described in Le F\`evre et al. (1995). The two galaxies common to  2  different masks were treated separately by averaging the flat-fielded and sky corrected 2D spectra obtained from each 2 mask. The $3\sigma$ detection limit in the shortest integration spectra is $\sim1.5\times10^{-18}$ ergcm$^{-2}$sec$^{-1}$\AA$^{-1}$ ~around 5000\AA, and our spectra would have allowed us to identify any emission line with a flux larger than $3 \times 10^{-17}$ ergcm$^{-2}$sec$^{-1}$. A total of 51 galaxies have secure redshift measurements in the range $0 \leq z \leq 1.3$; 2 stars were identified, and we failed to measure redshifts for 45 galaxies.

\section{Results}

McCarthy et al. (1990) noted that $Ly\alpha$ was the only feature in support of their redshift for MRC0316-257, while Eales et al. (1993) identified [OIII]5007 from IR spectroscopy. Our spectra  confirm $Ly\alpha$ in emission, as well as CIV1549\AA, HeII1640\AA ~and CIII]1909\AA ~(figure 1). The redshift of the radio-galaxy MRC0316-257 is therefore $z=3.1420\pm0.0020$. The V image shows another galaxy of roughly equal magnitude 2.7 arcseconds away, while in the 5007\AA ~image the eastern galaxy disappears almost completely. 

In addition to the radio-galaxy, we have identified spectroscopically two galaxies with a redshift similar to the radio-galaxy. The first one, ``galaxy A'', is 44 arcseconds away from the radio-galaxy. Its spectrum shows a strong line at 5030.3\AA, and faint features with marginal S/N at the same redshift (fig.2). A cross--correlation of the spectrum of galaxy A, after subtraction of the 5030\AA ~emission line, with a template spectrum with line widths taken from \cite{yeet} (1996), gives a marginal correlation at z=3.14 with a correlation coefficient R=2.8.  The  observed equivalent width of the emission line is 230$\pm60$\AA, which, if it was [OII]3727\AA ~at z=0.350, would indicate a rest frame $W_{OII}=170\pm44$\AA, larger than any  equivalent width for galaxies in the Canada-France Redshift Survey (CFRS), or 3 times larger than for any galaxies at $z\leq0.4$ in the  CFRS (\cite{ham}). Moreover,  all galaxies in the CFRS with rest $W_{OII}>30$\AA ~have rest $W_{OIII}>5$\AA ~and $W_{H\alpha}>40$\AA, which would have been easily identified in our spectra at a redshift $z=0.349$.  The identification of the line at 5030\AA ~with $Ly\alpha$ at $z=3.1378\pm0.0028$ seems therefore to be secure.

The second galaxy, ``galaxy B'', is located 182 arcseconds from MRC0316-257. Its spectrum shows $Ly\alpha$ and CIV1549\AA, and therefore gives a secure redshift $z=3.1351\pm0.0028$. Both CIV1549\AA, and to a lesser extent $Ly\alpha$ exhibit a broad component, which indicates the presence of an AGN.
 
Close examination of the images indicate that both MRC0316-257 and ``galaxy A'' are resolved under FWHM=1.5 arcseconds seeing, while ``galaxy B'' is only marginally resolved. After deconvolution from the observed PSF, the half-light radii of MRC0316-257 and ``galaxy A'', are $8.9\pm2.0$ and $11.6\pm1.1$ $h^{-1}$ kpc respectively. The fact that galaxy B is quite compact supports  the contribution of an AGN. The  V, I and $Ly\alpha$ images of galaxy A are strinkingly different: the peaks of continuum V and line emission are offset by 0.6 arcsec, with the peak of $Ly\alpha$ emission extending toward the SW, where little or no continuum light is detected in the V and I bands, and little or no emission is observed in I at the peak of $Ly\alpha$ emission, while several blobs of emission appear 3 arcseconds to the NW in the I band.

\section{Discussion}


In figure 3 we plot the (V$-$I) vs (V$-$5007\AA) color-color diagram. Besides  MRC0316-257, and the two associated galaxies identified in spectroscopy, two galaxies have significant excess emission in the 5007\AA ~filter (V$-$5007$>-$1.2). One of these galaxies has a redshift $z=0.335$, and the excess emission in the 5007\AA ~filter is coming from the [OII]3727\AA ~emission line. The other galaxy, galaxy C, has not been observed spectroscopically. Therefore, in our imaging 
field of view of 78.2 arcmin$^2$, we have three confirmed galaxies  at $z\sim3.14$, above our detection thresholds of $W_{Ly\alpha} > 12$\AA\ and $V < 23.8$, and a fourth galaxy, galaxy C, which remains a good but unconfirmed candidate galaxy at $z\sim3.14$.

Any estimate of the space density of $z \approx 3.14$ galaxies from our
observations is necessarily quite uncertain due to the small number of objects
which we detect.     
The maximum redshift depth probed is set by the range over which our narrow 
band filter would detect the Ly$\alpha$ emission line, i.e. $3.081 < z < 3.160$.
Under that assumption, the resulting co-moving cosmological volume surveyed
is therefore $13900 h_{50}^{-3}$Mpc$^3$ assuming $q_0 = 0.5$.  (For $q_0 = 0$,
the effective volumes are 7.4$\times$ larger, and the densities correspondingly
smaller.)
However, the three galaxies which we have confirmed spectroscopically are 
confined to a much smaller redshift range, close to the radio galaxy:  
$3.1351 < z < 3.1420$.  This corresponds to a rest--frame velocity difference 
$\Delta v = 500$km~s$^{-1}$, quite in line with expectations for galaxy 
clusters and/or ``walls'' of large--scale struture.   It is therefore likely 
that the effective ``cluster'' volume is much smaller than the value given 
above.    Assuming $\Delta z = 0.0069$, the corresponding co-moving volume 
would then be $1210 h_{50}^{-3}$Mpc$^3$.   We therefore may bracket the range
of possible space densities by assuming $3/13900 < n $ (Mpc$^{-3}) < 4/1210$,
or $2.2\times10^{-4} < n (h_{50}^3 $Mpc$^{-3}) < 3.3\times10^{-3}$,
with a ``best guess'' value of $3/1210 = 2.5\times10^{-3} h_{50}^3$Mpc$^{-3}$.

The population of $3.0 < z < 3.5$ field galaxies identified by 
Steidel {\it et al.} 1996 has a comoving space density of 
$3.6 \times 10^{-4} h_{50}^3$Mpc$^{-3}$ for $q_0$=0.5.   At face value,
this would imply that our ``best guess'' space density represents
an overdensity of $\sim 7\times$ compared to the field population at
similar redshift.  However, it is difficult to make a direct comparison 
between our ``cluster'' space density and the``field'' value since our 
galaxies 
were selected on the basis of strong Ly$\alpha$ flux, and have brighter 
continuum magnitudes ($V < 23.8$) than the Steidel {\it et al.} objects 
(which have $\cal{R}$ $< 25.5$).    Therefore the estimate given above for 
the cluster overdensity is almost certainly a substantial {\it underestimate.} 
Even our derived lower limit of $n > 2.2\times10^{-4} h_{50}^3 $Mpc$^{-3}$ 
for the 0316-257 field is substantially higher than the corresponding 
space density of field galaxies sharing the same continuum magnitudes and 
Ly$\alpha$ properties.  

The rest frame $Ly\alpha$ equivalent widths are $500\pm150$, $55\pm14$, $110\pm18$\AA, for the radio galaxy, and galaxies A and B respectively. While the radio-galaxy and galaxy B have resolved $Ly\alpha$, with $FWHM=30\pm3$\AA ~and $22\pm4$\AA ~respectively, and CIV is resolved for galaxy B, with $FWHM(CIV)=38\pm8$\AA ~after deconvolution by the instrumental response, $Ly\alpha$ is unresolved at our resolution for galaxy A. This may indicate that $Ly\alpha$ in galaxy A is produced mainly by stellar photo-ionisation.  Even after correcting for the contribution of the Lyman alpha line to the V band, Galaxy A is extremely blue, with  $V - I < -0.5$.  This is
very unusual for faint galaxies, and indeed Galaxy A is by far the
bluest object in our field of view, as can be seen from figure 3.
Galaxy A is bluer than most or all of the z$\sim$3 objects reported
by Steidel et al. (1996). It is also interesting that its Lyman alpha 
emission is significantly stronger than that in most of the Steidel
et al. Lyman break galaxies.  The blue color and strong Ly-a emission
may indicate an object dominated by OB-stars in a strong starburst
with very little dust. The complex morphology of galaxy A is quite different from the morphology of galaxies identified on the basis of the 912\AA ~discontinuity (\cite{stei}, \cite{gia2}), with complex spatial distribution of continuum and line emission  as described in section 3. While the evidence for old stars will have to be searched for at redder wavelength, this galaxy might well be forming its first stars and fit the definition of a proto-galaxy.

\section{Conclusion}

We have spectroscopically identified 2 galaxies at the same redshift as the radio-galaxy MRC0316-257, z$\sim3.14$. These galaxies exhibit $Ly\alpha$ in emission with $W(Ly\alpha)>50$\AA. One galaxy is resolved with a size $11 h^{-1}$ Mpc, has unresolved $Ly\alpha$, and extremely blue $V-I$ color, which might indicate that the main photo-ionising process is coming from a first generation of young stars in a low dust medium. The other galaxy is unresolved, and has broad $Ly\alpha$ and CIV emission lines, indicating the presence of an AGN. 

These observations of galaxies around MRC0316-257 provide tentative evidence for a region of high galaxy density, possibly indicative of clustering of galaxies occuring at $z=3.14$. They indicate that the search for galaxies around very high redshift radio-galaxies is indeed possible from $Ly\alpha$ imaging techniques.  However, the total number of galaxies identified with an observed  $Ly\alpha$ luminosity $>10^{-16}$ ergcm$^{-2}$sec$^{-1}$ is small, indicating the limits of this technique, in agreement with previous searches. Lyman break techniques, as successfuly applied by Steidel et al. (1996), will be  required to identify additional galaxies at the same redshift and confirm the existence of a cluster or proto-cluster of galaxies.

\acknowledgments

We would like to thank S. Charlot and H. Yee for useful discussions, the refereee, P. McCarthy, for his useful report, and the CFHT staff for their support during the observations.

%
%

\clearpage

\begin{deluxetable}{lccccccc}
\footnotesize
\tablewidth{7.2in}
\tablecaption{Properties of the galaxies at z$\simeq$3.14}
\tablehead{
\colhead{Object}                 & \colhead{$\alpha_{1950}$}        &
\colhead{$\delta_{1950}$}        & \colhead{$f_{Ly\alpha}$}& 
\colhead{V}                      & \colhead{I}                      & 
\colhead{$M_B^{(a,b)}$}          & \colhead{FWHM}\\ 
& & & (ergs$^{-1}$cm$^{-2}$) & & & &  (kpc)$^{(a)}$
}
\startdata
0316-257 & $03^h16^m02.^s66$ $^{(c)}$ & $-25^{\circ}46'04''$ $^{(c)}$  & 3.3$\pm$0.1 10$^{-16}$ & 23.24$\pm$0.04 & 22.53$\pm$0.06 & $-$23.27 & 8.9$\pm2.0$ \nl
A & $03^h15^m59.^s62$ & $-25^{\circ}45'52''$  & 1$\pm$0.1 10$^{-16}$ & 23.84$\pm$0.04 & 24.97$\pm$0.30 & $-$22.67 & 11.6$\pm1.1$ \nl
B & $03^h15^m50.^s32$ & $-25^{\circ}44'53''$ & 2.3$\pm$0.2 10$^{-16}$ & 23.21$\pm$0.03 & 22.67$\pm$0.09 & $-$23.30 & 2.3 $^{(d)}$ \nl
C $^{(e)}$ & $03^h15^m51.^s92$ & $-25^{\circ}42'38''$ & 0.5 10$^{-16}$ $^{(f)}$ & 23.87$\pm$0.06 & 23.48$\pm$0.10 & $-$22.64 & 6.3$\pm0.9$ \nl
\tablecomments{$^{(a)}$ Assuming $H_0=50$ kms$^{-1}$Mpc$^{-1}$ and $q_0=0.5$\\
$^{(b)}$ The K-correction has been estimated from Bruzual 1983\\
$^{(c)}$ From McCarthy et al. 1990\\
$^{(d)}$ Unresolved\\
$^{(e)}$ Candidate z=3.14 galaxy on the basis of narrow band imaging only\\
$^{(f)}$ Estimated from narrow band imaging}
\enddata
\end{deluxetable}

\normalsize

\clearpage

%
%

\figcaption[fig_imV_Lya.eps]{V image {\it (top)} and 5007\AA ~image {\it (bottom)} of a field $2\farcm15 \times 4\farcm$ around the radio-galaxy MRC0316-257 (RG). North is at the top, East to the left. Galaxies A and B clearly exhibit excess emission in the 5007\AA ~filter, which contains Ly$\alpha$ redshifted to $z\sim3.14$.}

\figcaption[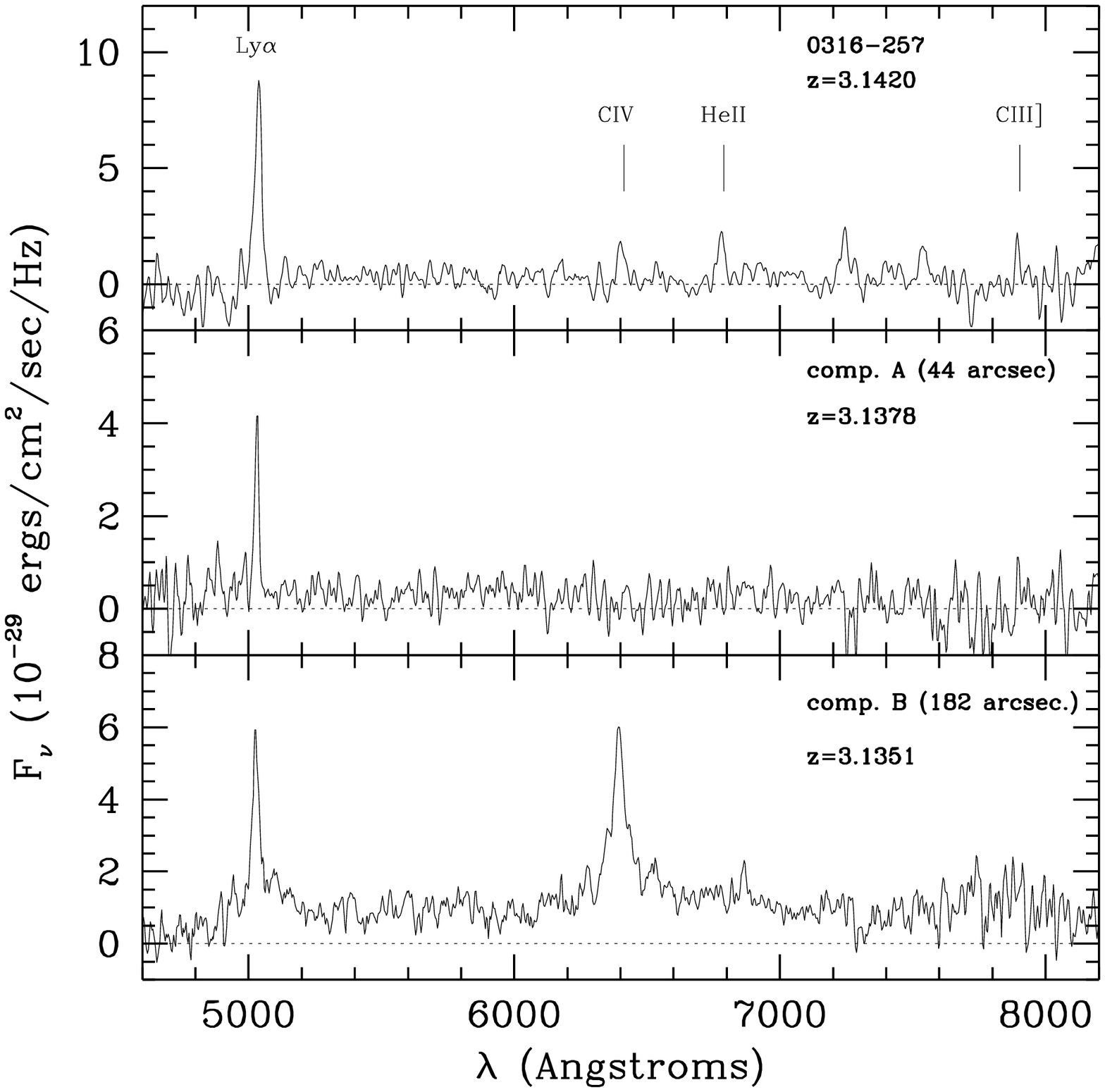]{CFHT spectra: radio-galaxy MRC0316-257 {\it (top)}, galaxy A {\it (middle)}, galaxy B {\it (bottom)}}

\figcaption[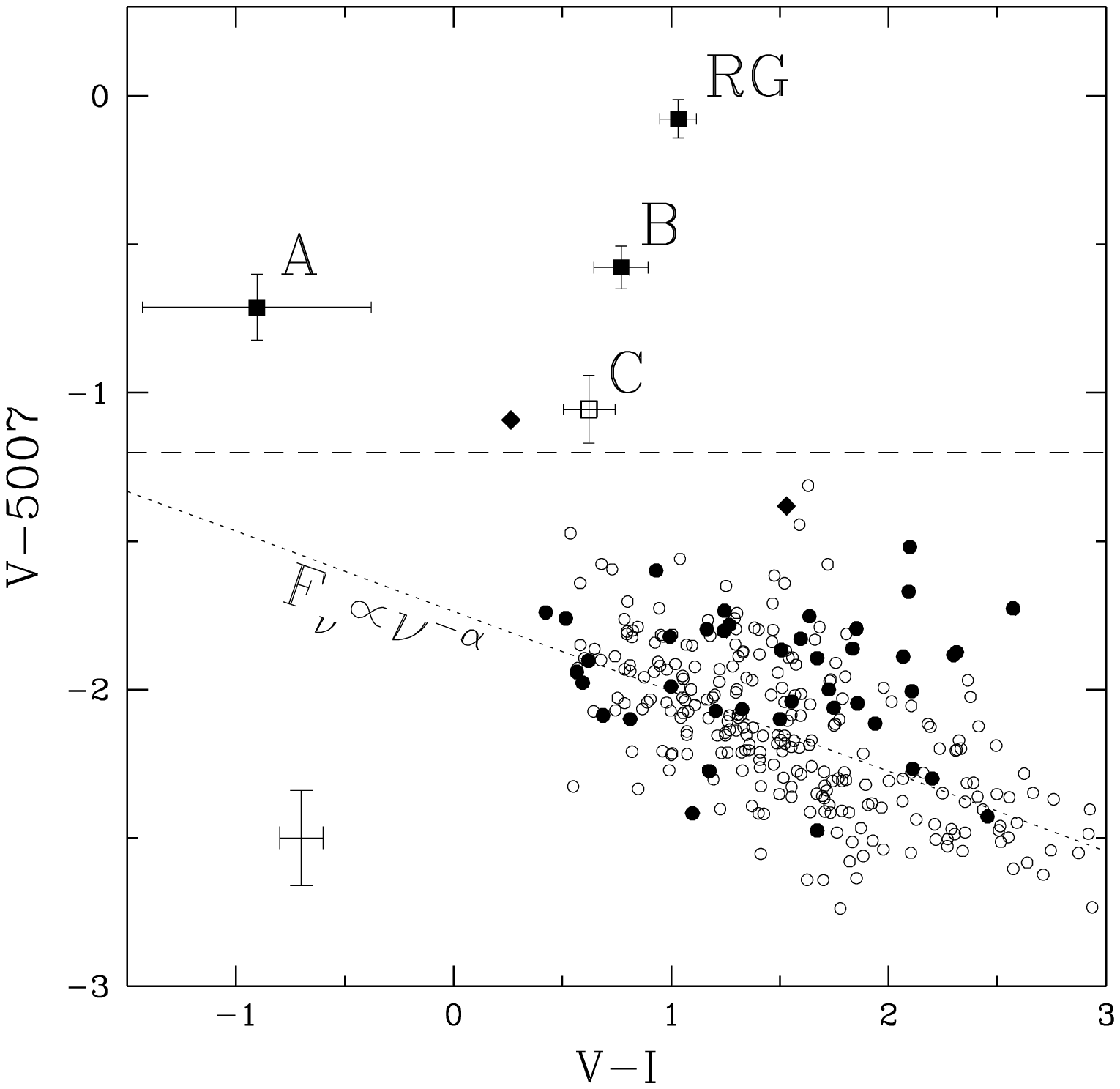]{(V$-$I) vs. (V$-$5007\AA) diagram. Solid symbols indicate spectroscopically observed galaxies: squares for $z\sim3.14$ galaxies, diamonds for the galaxies with $z\sim0.35$, circles for other galaxies with measured redshift. Open symbols indicate galaxies not observed spectroscopically: square for a galaxy with V$-$5007\AA$>-$1.2, a good candidate $z\sim3.14$ galaxy, circle for other galaxies. Note that the V-I for galaxy A is $V-I=-0.7$ after correction for the $Ly\alpha$ line contamination in the filter bandpass,  still extremely blue.}





\begin{thebibliography}{}
\bibitem[Bruzual 1983]{bru} Bruzual, A.G., 1983, ApJS, 53, 497
\bibitem[Dickinson 1994]{dic} Dickinson, M., 1994, PhD thesis
\bibitem[Dickinson 1995]{dic2} Dickinson, M., 1995, in ``Galaxies in the Young Universe'', eds. H. Hippelein, K. Meisenheimer, \& H-J. R\"oser, Springer, p.144.
\bibitem[Eales\&Rawlings 1993]{eales} Eales, S.A., \& Rawlings, S., 1993, \apj, 411, 67
\bibitem[Evrard\&Charlot 1994]{evr} Evrard, A.E., \& Charlot, S., 1994, \apj, 424, L14
\bibitem[Fontana et al. 1996]{fon} Fontana, A., Cristiani, S., D'Odorico, S., Giallongo, E., Savaglio, S., 1996, \mnras, 279, L27
\bibitem[Francis et al. 1996]{fran} Francis, P.J., Woodgate, B.E., Warren, S.J., Moller, P., Mazzolini, M., Bunker, A.J., Lowenthal, J.D., Williams, T.B., Minezaki, T., Kobayashi, Y., Yoshii, Y., 1996, \apj, 457, 490
\bibitem[Frenk et al. 1996]{frenk} Frenk, C.S., Evrard, A.E., White, S.D.M., Summers, F.J., 1996, \apj, preprint
\bibitem[Giavalisco et al. 1994]{gia}  Giavalisco, M., Steidel, C.C., \& Szalay, A.S., 1994, \apj 425, L5
\bibitem[Giavalisco et al. 1996]{gia2} Giavalisco, M., Steidel, C.C., \& Macchetto, F., 1996, \apjl, in press
\bibitem[Hammer \& al. 1996]{ham} Hammer, F., Flores, H., Lilly, S.J., Crampton, D., Le F\`evre, O., Rola, C., Mallen-Ornelas, G., Schade, D., Tresse, L., \apj, preprint
\bibitem[Hill\&Lilly 1991]{hill} Hill, G.J., Lilly, S.J., 1991, \apj, 367, 1
\bibitem[Le F\`evre et al. 1994a]{olf1} Le F\`evre, O., Crampton, D., Hammer, F., Lilly, S.J., Tresse, L., 1994a, \apjl, 424, L14
\bibitem[Le F\`evre et al. 1994b]{olf2} Le F\`evre, O., Crampton, D., Felenbok, P., Monnet, G., 1994b, \aap, 282, 325
\bibitem[Le F\`evre et al. 1995]{olf3} Le F\`evre, O., Crampton, D., Lilly, S.J., Hammer, F., Tresse, L., 1995, \apj, 455, 60
\bibitem[McCarthy et al. 1990]{Mcc} McCarthy, P.J., Kapahi, V.K., van Breugel, W., Subrahmanya, C.R., 1990, \aj, 100, 1014
\bibitem[Pascarelle et al. 1996]{pas} Pascarelle, S.M., Windhorst, R.A., Driver, S.P., Ostander, E.J., \& Keel, W.C., 1996, \apjl, 456, L21
\bibitem[Peebles et al. 1989]{pee} Peebles, P.J.E., Daly, R., Juszkiewicz, 1989, \apj, 347, 563
\bibitem[Steidel et al. 1996]{stei} Steidel, C.C., Giavalisco, M., Pettini, M., Dickinson, M., Adelberger, K., 1996, \apjl, in press
\bibitem[Yee\&Green 1984]{yee} Yee, H., Green, 1984, \apj, 280, 79
\bibitem[Yee et al.]{yeet} Yee, H., Ellingson, E., Bechtold, J., Carlberg, R., Cuillandre, J.C., 1996, AJ, in press (May 1996)
\end{thebibliography}
\end{document}